# Designing Ethical Learning for Agentic AI: Toegye Yi Hwang's Ethical Emotion Regulation Framework

Ji Yeon Kim

*Abstract*—**Agentic AI systems capable of autonomous goal setting and proactive intervention introduce new challenges for regulating moral-emotional processes in learning environments. Existing frameworks typically treat emotion as reactive feedback or engagement optimization, overlooking the need for normative regulation across autonomous decision cycles. This paper proposes an ethical emotion regulation framework for agentic AI learning design inspired by Toegye Yi Hwang's moral-emotional philosophy. The Ethical Emotion Feedback System (EEFS) is reconstructed as a five-stage architecture aligned with agentic cycles, articulating stage-specific design principles, scenario classifications, and an EEFS Evaluation Instrument to enable systematic assessment of moral-emotional alignment in agentic AI systems.**

## I. INTRODUCTION

Agentic AI systems are becoming increasingly embedded in educational technologies. Unlike reactive systems that rely on user-initiated inputs, they enable autonomous decision-making, goal-directed behavior, and proactive intervention in learning processes [1], [2], [3]. These systems perceive learner states, plan and execute interventions, evaluate outcomes, and adapt over time [4]. While such autonomy enhances personalization, it also raises ethical concerns regarding how AI–learner emotional interactions are structured and regulated.

Dieterle et al. [5] identify five ethical divides in AI-enhanced education—access, representation, algorithmic bias, interpretation, and citizenship—emphasizing that these challenges are systemic and cyclical, with inequalities in one domain reinforcing others. From this perspective, ethical principles are more effective when embedded in system architecture rather than imposed as external constraints. In agentic systems, where AI autonomously initiates and sustains interaction, emotional responses can amplify bias, underscoring the need to regulate emotion at the architectural level.

Research further underscores the importance of emotional engagement in AI-mediated learning. A recent review of the field suggests that chatbot-supported learning environments can shape students' emotional engagement [2]. These dynamics are likely to intensify in agentic systems, where AI autonomously governs the timing, content, and tone of interaction. Kostopoulos et al. [2] extend this discussion to agentic educational contexts, suggesting that when emotional exchanges are autonomously initiated, emotional regulation shifts from a peripheral interactional feature to a structural design consideration.

However, existing scholarship does not sufficiently conceptualize emotional interaction as a distinct ethical domain. A comprehensive 2019 review of global AI ethics guidelines identified convergence around a limited set of high-level principles, including transparency, justice, and accountability [6]. Yet emotional interaction was not articulated as an independent ethical domain, and this gap remains largely unaddressed in subsequent frameworks. Reviews of multimodal AGI similarly articulate ethics at a high level of generality [7]. At the technical level, while systems such as SMES demonstrate advances in supportive response sequencing within therapeutic settings [8], these approaches remain task-bounded and do not extend to stage-level ethical self-regulation within agentic educational systems. These converging limitations reveal a structural absence of normative architectures capable of governing emotional processes within agentic systems.

In response to this gap, virtue ethics offers an alternative by grounding moral conduct in relational dispositions and cultivated character rather than fixed rules. Within this orientation, Confucian ethics has drawn attention for its focus on relational context, emotional cultivation, and moral responsiveness. Seok [9] argues that Confucian virtue ethics provides a model of moral reasoning that complements deontological and consequentialist traditions dominant in AI ethics discourse. Wong [10] situates Confucian technology ethics within a framework grounded in Dao, harmony, and personhood, extending this paradigm to socio-technical systems. Lee et al. [11] further develop this line of work by proposing a triadic model integrating Confucian fiduciary ethics, intelligence augmentation, and agentic AI design. However, these contributions remain primarily conceptual and do not specify stage-level mechanisms through which ethical principles regulate emotional processes within autonomous decision cycles.

Among the Confucian concepts most directly relevant to this operationalization challenge is Toegye Yi Hwang's dual-aspect account of moral-emotional arousal. His framework provides a normative differentiation between principle-

---

Ji Yeon Kim is with Seoul National University, Seoul, South Korea (e-mail: jiyeonkim.net@snu.ac.kr).

originated and stimulus-originated emotions, a distinction that process-oriented regulation models typically do not capture. To address this gap, this study extends the Ethical Emotion Feedback System (EEFS), a five-stage model originally proposed by Han [12] and grounded in this Toegye-derived framework [13], [14]. The present study reconstructs EEFS as a regulatory architecture embedded within the agentic decision cycle, operationalizing Toegye's philosophical distinctions as stage-level design mechanisms.

Although inspired by Neo-Confucian moral-emotional philosophy, EEFS is presented here as a general design framework for ethical emotional regulation in agentic AI learning systems. Self-reflective monitoring, relational calibration, and long-term ethical equilibrium are treated as structural requirements of autonomous systems engaging learners over time.

This study develops a regulatory architecture for ethical emotional regulation in agentic AI learning systems. It reconstructs the Ethical Emotion Feedback System (EEFS) as a stage-level framework embedded in the autonomous decision cycle. EEFS is translated into operational design principles and interaction scenarios, accompanied by an evaluation instrument to assess ethical emotional regulation at the design and prototyping stages. This approach operationalizes philosophical design principles at the level of system architecture.

## II. THEORETICAL BACKGROUND

### A. Agentic AI in Education

Agentic AI refers to systems capable of autonomous decision-making, goal-directed behavior, and proactive intervention [1], [2], [3], extended through capabilities such as planning, reflection, tool use, memory, and multi-agent collaboration [1], [15], [3]. Unlike prior generations of AI tools that respond to user inputs, these systems can independently determine when and how to intervene in learning processes. This distinction has direct consequences for emotional interaction design.

Drawing on prior characterizations of agentic capabilities [1] and operational patterns in educational contexts [2], this study conceptualizes the autonomous cycle of agentic AI as a six-stage process consisting of PERCEIVE, REASON, PLAN, ACT, EVALUATE, and LEARN. The cycle operates continuously, with the agent initiating interactions based on its own perception and reasoning. This capacity raises ethical stakes that reactive AI tools do not pose: when AI determines the timing, scope, and emotional tone of engagement, learners lose the opportunity to set boundaries in advance.

In this context, agentic AI raises questions regarding learner autonomy, emotional safety, and the boundaries of AI agency. One recent study suggests that agentic AI can support self-efficacy and motivational autonomy in higher education [16]. However, the emotional and ethical architecture governing how such support is initiated, calibrated, and sustained remains underspecified.

### B. Emotional Aspects in Agentic AI for Learning

Informed by recent studies of agentic AI and emotion in education, this study identifies four core emotional functions for learning environments: Emotion Perception, Emotion Expression, Emotion Regulation, and Emotional Relationship.

Emotion Perception refers to the AI's ability to infer learner affect through multimodal signals such as facial, vocal, textual, and physiological cues [17]. Recognizing confusion, frustration, or disengagement is critical for sustaining learner motivation [2], [8]. However, emotional states cannot be reliably inferred from single modalities, and expression varies across contexts and cultures [18]. Emotion data constitutes a sensitive category of personal information, requiring heightened ethical safeguards due to risks of uncertainty and misclassification [19]. Meta-analytic evidence indicates moderate effects of emotional AI interventions, with substantial variability depending on modality and support type [20]. A complementary systematic review of AI-based emotion integration in learning environments corroborates this variability across contexts [21]. In agentic systems, emotion perception functions as the input layer of the autonomous decision cycle and must therefore be designed with epistemic caution and robustness safeguards.

Emotion Expression concerns how AI manifests affect in interaction. Studies show that affective pedagogical agents can enhance learner motivation and trust through expressive cues such as facial behavior, posture, and vocal tone [22]. For agentic AI, expression becomes particularly consequential because the system may autonomously initiate emotional exchanges independent of learner solicitation. Design principles for empathic conversational agents emphasize calibrated, context-sensitive, and personalized emotional feedback [23].

Emotion regulation concerns how AI modulates its emotional responses to sustain appropriate interaction patterns. Existing approaches primarily frame regulation as learner engagement optimization, leaving a gap between performance-oriented models and the kind of normative self-regulation that ethical emotional conduct requires. Findings from AI companion use outside formal educational contexts suggest that AI-mediated emotional support can facilitate self-reflection and regulation, although with mixed implications for social relationships and help-seeking behaviors [24].

Emotional Relationship encompasses the development and maintenance of trust and relational dynamics over sustained interaction. Users may experience AI interactions as relationships rather than transactions [25]. For agentic systems operating across multiple sessions and contexts, long-term socioaffective alignment becomes critical, as interaction patterns co-evolve within a shared social and psychological ecosystem [25].

### C. Toegye Yi Hwang's Emotional Ethics and EEFS

Toegye Yi Hwang (퇴계 이황; 退溪 李滉, 1501–1570), a Korean Neo-Confucian philosopher comparable to John Dewey in philosophical significance, was a leading Korean Neo-Confucian scholar whose work on moral emotions made a distinctive contribution to East Asian ethics [13]. His Four–Seven Debate on Emotions (사단칠정론, 四端七情論) examines the relationship between moral emotions, such as compassion, shame, modesty, and discernment, and general emotions, such as joy, anger, sorrow, and fear. He argued that the former originate in Principle $i$ (이, 理), whereas the latter are associated primarily with material force $gi$ (기, 氣) and require attentive cultivation to maintain proper measure. This relation was formalized in his doctrine of *Ibalgisu* (이발기수, 理發氣隨), meaning "Principle issues, and material force follows" [12], [14].

Within this framework, Toegye articulated a dual-aspect account of emotional arousal [12], [14]. The Four Emotions, rooted in Principle, constitute the primary normative foundation of virtue: as innate moral dispositions, they are inherently good and serve as the basis for ethical action. The Seven Emotions, associated with material force, function as the regulatory dimension of moral life: as ordinary emotional responses to external stimuli, they are not inherently evil but become morally distorted when material force exceeds due measure and obscures Principle. Self-cultivation thus requires both nourishing the Four as one's primary moral orientation and regulating the Seven to prevent their excesses from disrupting moral balance. This dual-aspect model of moral cultivation provides the conceptual foundation for the EEFS stage sequence developed below.

Central to Toegye's framework for ethical emotional regulation is the concept of Gyeong (경, 敬; reverence), an attitude of attentive awareness and reflective monitoring that supports the regulation of emotional responses. Toegye positioned Gyeong not as a discrete evaluative act but as an integrative practice unifying self-reflection (*jaseong* and mind cultivation, governing both pre-arousal and post-arousal states of the mind, which he described as the "beginning and end of sagely learning" [26], [27]. This persistent, non-punctual character of *Gyeong* is precisely what makes it translatable as a metacognitive loop rather than a single checkpoint in AI system design. Complementing this internal orientation, *Ye* (예, 禮; ritual propriety) provides a relational structure for expressing emotions appropriately within social contexts. Together, these concepts form a coherent regulatory model in which emotions function both as authentic moral signals and as context-sensitive responses calibrated to relational norms [12], [13], [26].

This parallels Dewey's insight that "experience is emotional but there are no separate things called emotions in it" [28]. In both accounts, emotion is not an isolable entity but a dimension permeating experience as a whole. Harroff [29] interprets this parallel as evidence of the cross-cultural translatability of Toegye's relational emotional philosophy. By integrating metaphysical grounding, dual-aspect emotional structure, reflective cultivation, and relational calibration into a coherent normative system, Toegye's framework becomes particularly suited to agentic AI systems. Rather than relying solely on externally imposed constraints, such systems require internally structured processes of ethical emotional regulation, a requirement that aligns with Toegye's account of cultivation.

Building on this framework, the EEFS model operationalizes this conceptual model into a five-stage cyclical architecture for AI emotional ethics [12]. The stages are functionally defined as follows: (1) *Ibal* (이발, 理發; Ethical Principle Activation), activation of ethical constraints prior to emotional response generation; (2) *Gisu* (기수, 氣隨; Response Generation following Principle), production of emotionally expressive output under normative guidance; (3) *Gyeong* (경, 敬; Mindful Awareness, Self-Reflective Loop), internal monitoring to assess the ethical appropriateness; (4) *Ye* (예, 禮; Relational Calibration), contextual adjustment of emotional expression to avoid excess or deficiency in social interaction; and (5) *Junghwa* (중화, 中和; Ethical Equilibrium), stabilization of system behavior through iterative cycles of activation, generation, monitoring, and calibration, corresponding to sustained ethical alignment [12].

The original EEFS formulation proposed directional principles for AI emotional ethics across healthcare, educational, and social robotics applications [12]. The present study narrows this scope to agentic AI learning systems and deepens the framework's theoretical grounding to support concrete design specifications.

*D. Design Gap Clarification*

Despite growing attention to ethical AI and emotional AI in educational contexts, existing models remain limited in translating ethical principles into actionable design mechanisms. Although the IEEE Standard for Ethical Considerations in Emulated Empathy establishes requirements for ongoing monitoring and quality assurance in systems implementing emulated empathy [30], these provisions primarily address general autonomous systems and do not explicitly engage with the recursive decision architectures that characterize agentic learning agents. Prior frameworks articulate ethical principles or procedural models for emotional support; however, they stop short of systematically embedding ethical regulation within the autonomous decision cycles of agentic systems.

AIEd systems are typically organized around three interacting models: pedagogical, domain, and learner [31]. Learner models track activities, achievements, and learner state information to determine subsequent interactions. However, within this architecture, emotional information is treated primarily as input for adaptive optimization rather than as a domain subject to ethical regulation. Emotion is thus operationalized as a variable for engagement or performance enhancement, rather than as a dimension requiring normative structuring within the decision process itself.

More specifically, two critical design-level mechanisms remain underdeveloped. First, current models lack a structured self-reflective loop through which an AI system can evaluate the ethical alignment of its emotional responses before or during decision execution. Emotional regulation is typically framed as an optimization problem oriented toward learner engagement, rather than as an embedded process of ethical self-assessment within system architecture. Second, existing approaches do not provide a temporal equilibrium mechanism capable of sustaining ethical consistency across extended interactions. Agentic systems in learning environments operate across multiple sessions and evolving relational contexts. In the absence of a mechanism that stabilizes ethical orientation over time, emotional conduct risks gradual drift toward inconsistency or misalignment with pedagogical values. As a result, neither existing AI ethics frameworks nor emotional AI models such as SMES [8] were designed to address how ethical principles should regulate emotion across the autonomous decision cycles characteristic of agentic learning systems.

This paper addresses these design gaps through EEFS's five-stage regulatory architecture embedded in agentic decision cycles. Although agentic systems include a perceptual input layer (PERCEIVE), it does not constitute an independent ethical stage. Normative interpretation begins at Ethical Principle Activation (Ibal), where contextual inputs are aligned with principled constraints before response generation. EEFS thus reframes emotion not as a mere engagement variable but as a normatively structured component of autonomous decision architecture.

## III. EEFS FRAMEWORK FOR AGENTIC AI LEARNING DESIGN

### A. The Agentic AI Autonomous Cycle

The autonomous cycle of agentic AI comprises six stages that operate continuously in learning environments. In the PERCEIVE stage, the system monitors the learning environment, including learner behavioral patterns, performance trajectories, and emotional states inferred through multimodal cues. The REASON stage involves goal-setting and decision-making about whether and how to intervene, drawing on the system's model of learner needs and educational objectives. During the PLAN stage, the system formulates intervention strategies, including content selection, timing, and emotional tone. The ACT stage executes the planned intervention, which may involve proactive outreach, feedback delivery, or learning content presentation. In the EVALUATE stage, the system assesses intervention outcomes and learner responses. Finally, the LEARN stage updates the system's memory and models based on accumulated experience.

Unlike reactive systems that wait for learner input, agentic AI may initiate the decision cycle at any point based on its autonomous perception and reasoning. This capacity raises the ethical stakes of emotional interaction. The learner may not have initiated or anticipated the exchange and therefore has limited opportunity to define expectations or boundaries for the system's emotional conduct.

This capacity for autonomous initiation introduces two structural dimensions that differentiate agentic emotional interactions. The first concerns who initiates the interaction, whether learner-initiated or system-initiated. The second concerns the temporal scope of the interaction, whether episodic or longitudinal.

Yan's [15] conceptualization of agentic AI as a socio-cognitive teammate aligns with this distinction. The transition from passive tool to active participant alters the initiation dynamics of emotional exchange. These dimensions carry distinct ethical implications. The first concerns the legitimacy of unsolicited emotional engagement. The second concerns the maintenance of ethical consistency across sustained interactions. Together, they define the design space of interaction modes that the EEFS framework is intended to regulate.

### B. Architectural Mapping of the Agentic Cycle and EEFS

Table I maps the agentic decision cycle to emotional functions and EEFS stages, illustrating how ethical emotional regulation is embedded across the cycle. Figure 1 visualizes this integration within the overall decision process.

TABLE I
Architectural Mapping of the Agentic Decision Cycle and EEFS Stages

| Agentic Cycle | Emotional Function | EEFS Stage | Design Logic Label |
|---|---|---|---|
| PERCEIVE | Emotion Perception | (Input Layer) | — |
| REASON | Goal Setting | Ibal (이발, 理發; Ethical Principle Activation) | Design-Time Constraint |
| PLAN | Emotional Structuring | Gisu (기수, 氣隨; Response Generation following Principle) | System Behavior Layer |
| ACT | Emotion Expression | Ye (예, 禮; Relational Calibration) | Social Alignment Mechanism |
| EVALUATE | Emotion Regulation | Gyeong (경, 敬; Mindful Awareness, Self-Reflective Loop) | Meta-Cognitive Loop |
| LEARN | Emotional Relationship | Junghwa (중화, 中和; Ethical Equilibrium) | Temporal Ethical Equilibrium |

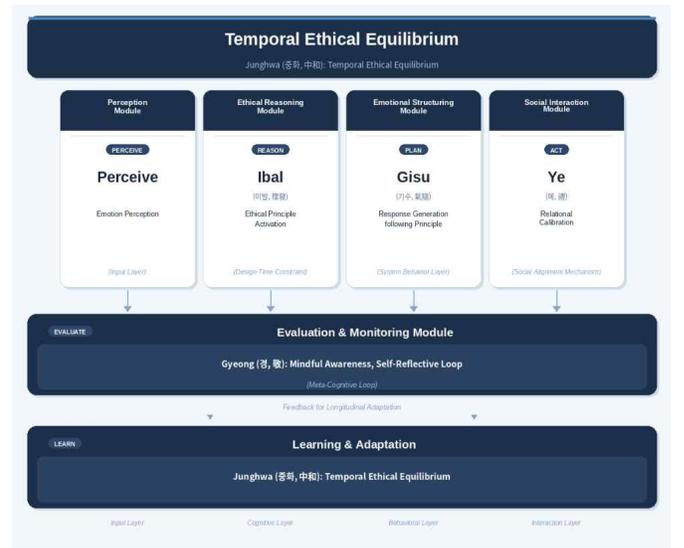

Fig. 1. Architectural mapping of EEFS stages onto the agentic AI learning system ((visualization prepared with assistance from Claude, based on the author's conceptual design).

This integration clarifies how EEFS structures ethical emotional regulation across the agentic decision cycle. At the PERCEIVE stage, emotion perception supplies emotional input but does not yet activate EEFS processing. When the system enters REASON, the Ibal stage imposes normative constraints that shape goal formulation. During PLAN and ACT, the Gisu and Ye stages regulate the generation and relational calibration of emotional responses. At EVALUATE, the Gyeong mechanism functions as an internal evaluative process that assesses alignment between emotional conduct and ethical principles before response finalization. Finally, the LEARN stage advances toward Junghwa, stabilizing long-term ethical orientation across iterative interactions.

*C. EEFS-based Learning Design Principles*

This section articulates design principles for each stage of the Ethical-Emotional Feedback System (EEFS), situating them within the system's architectural workflow. This stage-based structure shares a formal resemblance to Gross's process model of emotion regulation [32], which conceptualizes regulation as a temporally ordered sequence consisting of situation selection, situation modification, attentional deployment, cognitive change, and response modulation rather than as a single-point intervention [17], [23]. However, the similarity is structural rather than normative. Whereas Gross describes how emotions are modulated, EEFS specifies which modulations are ethically permissible and on what grounds, modeling ethical emotional regulation as a cyclical process in which each stage incrementally refines ethical alignment within learning interactions.

1) *Ibal — Design-Time Constraint:* Before initiating any goal-directed behavior, the system activates a set of ethical principles that constrain the permissible space of goals and interventions. These include dignity (safeguarding learner self-esteem and avoiding shame), fairness (ensuring equitable support regardless of learner characteristics), care (balancing emotional well-being with academic achievement), and responsibility (considering the long-term implications of interventions). At this stage, the system verifies whether each potential action upholds these ethical constraints to define the boundaries of acceptable behavior.

2) *Gisu — System Behavior Layer:* Once these ethical parameters are activated, the system generates emotional responses guided by them rather than optimized solely for engagement or task efficiency. Feedback should remain encouraging and authentic while avoiding extremes such as over-praising or emotional detachment. This layer transforms computationally generated responses into ethically regulated communicative behavior that promotes learning without inducing anxiety, shame, or artificial enthusiasm.

3) *Ye — Social Alignment Mechanism:* Emotional expressions are calibrated according to relational and cultural context, including learner background, interpersonal history, and the learner's current emotional state. A motivational message that benefits one learner may overwhelm another; therefore, this mechanism dynamically adjusts valence, intensity, and tone to preserve both contextual appropriateness and ethical integrity.

4) *Gyeong — Meta-Cognitive Loop:* Before finalizing an emotional response, the system performs meta-cognitive self-evaluation to assess alignment with ethical constraints. It examines potential risks of emotional harm, respect for learner autonomy, contextual appropriateness, and coherence with Ibal's ethical framework. If concerns are detected, the system regenerates or adjusts the response. This reflective loop operationalizes the monitoring requirements outlined in the IEEE Standard for Ethical Considerations in Emulated Empathy within the system's decision-making cycle.

5) *Junghwa — Temporal Ethical Equilibrium:* Through the continuous circulation of ethical activation, behavioral regulation, self-reflection, and contextual calibration, the system sustains long-term ethical equilibrium. Drawing on Toegye's conception of Deok (덕, 德) as dynamic moral balance, the system monitors longitudinal consistency across interactions, identifies potential drifts from ethical principles, and recalibrates its approach accordingly. This temporal feedback loop supports stability, trust, and ethical continuity in human–AI educational relationships.

*D. EEFS-Enhanced Learner Model*

Effective application of EEFS requires a learner model that captures not only cognitive and behavioral dimensions but also emotional and relational dimensions relevant to ethical emotional regulation. Building on the learner-centric design tradition in AIEd [33], learner models have progressively incorporated richer representations of learner interactions, including emotional information [31]. This study extends this tradition by introducing three ethically grounded domains: emotional, relational, and ethical history. Together with the cognitive domain, these form a four-domain EEFS-enhanced learner model oriented toward ethical regulation rather than engagement optimization alone.

The Cognitive Domain represents conventional learner modeling components such as knowledge level, misconceptions, learning style, and prior knowledge. The Emotional Domain, essential for EEFS application, includes current emotional state, temporal emotional patterns, frustration thresholds, anxiety levels, and motivational indicators. The Relational Domain captures the learner's evolving relationship with the AI system, including trust level, preferred interaction style, cultural context, and receptiveness to proactive intervention. The Ethical History Domain records prior ethically relevant interactions, including past interventions, learner responses, patterns of ethical success or concern, and longitudinal information informing the Junghwa stage's temporal equilibrium mechanism.

This four-domain model enables the EEFS stages to operate with contextually grounded information. The Ye stage draws on emotional and relational domains to calibrate responses, while the Junghwa stage leverages ethical history to sustain long-term ethical consistency. Collectively, these domains support the kind of rich learner models integrating cognitive, behavioral, and emotional signals identified by Kostopoulos et al. as essential for pedagogically aware agents [2].

# IV. EEFS APPLICATION SCENARIOS IN AGENTIC AI LEARNING DESIGN

## A. Scenario Classification Framework

The three scenarios are derived from the intersection of two dimensions grounded in agentic AI research and the EEFS architecture.

The first dimension concerns interaction initiation, that is, who triggers the emotional exchange. Prior work characterizes the shift from reactive to proactive AI as a defining feature of agentic systems, emphasizing autonomy and goal-directed behavior [1], [2]. This dimension yields two categories: learner-initiated (reactive) and system-initiated (proactive) interactions. The distinction carries ethical implications. When the AI initiates emotional engagement, the learner has not explicitly solicited the exchange, thereby increasing the system's ethical responsibility.

The second dimension concerns temporal scope, distinguishing episodic interactions (bounded within a single session) from longitudinal interactions (extending across multiple sessions over time). This distinction follows from two sources: the memory capabilities of agentic AI systems, which support persistent learner models across sessions [2], and the EEFS architecture itself, where the Junghwa stage addresses ethical equilibrium over extended interaction trajectories. Kirk et al. [25] describe this shift as a transition from transactional interaction to sustained socioaffective alignment.

Cross-tabulation of these dimensions initially yields four logical possibilities. However, in longitudinal contexts, the initiation distinction becomes less salient. As Kirk et al. [25] note, sustained human–AI interaction tends toward socioaffective co-evolution, where both parties alternate roles over time. Across extended relationships, learner–system pairs accumulate both reactive episodes, such as learner disclosures, and proactive episodes, such as system outreach.

Such longitudinal co-evolution shifts the primary ethical concern from episode-level appropriateness to trajectory-level coherence, namely whether trust, relational style, and ethical principles remain consistent across time. Distinguishing longitudinal interactions by initiation mode would therefore obscure this shared ethical focus. Consequently, the two longitudinal cells collapse into a single relational category, yielding three scenario types (see Table II).

TABLE II
Scenario Classification and Differential EEFS Stage Emphasis

| Scenario | Initiation Mode | Temporal Scope | Primary EEFS Stages |
|---|---|---|---|
| A. Reactive | Learner-initiated | Episodic | Gisu + Gyeong |
| B. Proactive | System-initiated | Episodic | Ibal + Ye |
| C. Relational | Both(accumulated) | Longitudinal | Junghwa + Ye |

Each scenario foregrounds a distinct ethical challenge and emphasizes specific EEFS stages as primary regulatory mechanisms. The Reactive Scenario focuses on the ethical quality of emotional response (Gisu) and its reflective evaluation (Gyeong). The Proactive Scenario emphasizes the legitimacy of autonomous initiation (Ibal) and the contextual calibration of outreach (Ye). The Relational Scenario addresses long-term ethical coherence (Junghwa) and the sustained calibration of emotional expression within an evolving relationship (Ye). Although all five EEFS stages operate across scenarios, this differentiated emphasis highlights the distinct regulatory priorities of each interaction mode while collectively demonstrating the full functional range of the framework.

## B. Reactive Scenario: Ethical Response to Learner-Initiated Emotional Disclosure

• Classification: Learner-initiated × Episodic.
• Primary ethical challenge: generating an emotionally appropriate and ethically regulated response within a single interaction episode.
• Situation: A learner has failed the same problem five times and types: "I'm terrible at math. I give up."
PERCEIVE: The system detects high frustration through text sentiment analysis and behavioral pattern (repeated failures). The emotional domain of the learner model indicates high test anxiety, underscoring the time-critical nature of intervention given the known frustration–disengagement trajectory [34].
REASON + Ibal: Ethical principles activate. Priority: learner dignity (protect self-esteem), care (emotional support). The goal shifts from problem-solving to emotional support.
PLAN + Gisu [Primary]: The system generates a regulated response that avoids dismissing the learner's emotional experience (e.g., generic reassurance) or reinforcing perceived inadequacy. Instead, it attributes difficulty to task demands rather than to learner deficiency and offers a choice of alternative approaches or a brief pause to preserve dignity while sustaining forward progress.
ACT + Ye: Calibrates to learner context. Given high anxiety, the system emphasizes choice and control rather than directive guidance.
EVALUATE + Gyeong [Primary]: Self-reflection evaluates whether the response risks inducing shame, offers agency without pressure, and acknowledges difficulty without labeling the learner, thereby performing ethical quality assurance before delivery.
LEARN + Junghwa: Records this interaction in the learner's ethical history and updates the learner model's emotional domain, contributing data for future relational processing.

## C. Proactive Scenario: Ethical Justification of Autonomous Intervention

• Classification: System-initiated × Episodic.
• Primary ethical challenge: establishing the legitimacy of autonomous outreach when the learner has not requested interaction.
• Situation: The agentic AI detects that a learner has not logged in for three days, recent quiz scores show a declining trajectory, and a major exam is scheduled next week.

PERCEIVE: The system monitors learning management data without learner initiation, identifying a risk pattern through behavioral absence and performance decline.

REASON + Ibal [Primary]: Before setting an intervention goal, ethical principles activate. This stage carries the primary ethical weight in the Proactive Scenario because the fundamental question is one of legitimacy: "Is proactive outreach justified? Does it respect learner autonomy?" The system must weigh the principle of care (potential learner benefit) against the principle of autonomy (the learner's right not to be contacted).

PLAN + Gisu: The system plans an outreach message. It avoids: "You haven't logged in—you should study" (judgmental, pressuring). It generates: "I noticed you have an exam coming up. If you'd like, I can help you review some topics. No pressure—just here if you need me." The message framing emphasizes availability rather than obligation.

ACT + Ye [Primary]: The system checks the relational context, including the learner's preference for email. Because the current time is evening, it delays the message until morning to avoid intrusion.

EVALUATE + Gyeong: Self-reflection asks: "Does this message pressure the learner? Does it respect their right to choose not to engage?" Confirms that the message offers support without obligation.

LEARN + Junghwa: Records the proactive outreach and tracks learner response to calibrate future contact thresholds.

### D. Relational Scenario: Long-Term Ethical Equilibrium in Sustained Interaction

• Classification: Both initiation modes × Longitudinal.

• Primary ethical challenge: maintaining ethical consistency and appropriate relational calibration across an extended series of interactions, where cumulative patterns supersede individual episodes.

• Situation: A learner has been using the system for three months. The Junghwa stage identifies a pattern: the learner responds positively to gentle humor but becomes defensive when the system is overly formal.

This scenario illustrates the temporal ethics dimension of EEFS. Unlike the episodic scenarios, which follow a single decision cycle, the Relational Scenario operates across accumulated cycles; accordingly, the following analysis foregrounds the EEFS stages that bear primary ethical weight in the longitudinal context.

Junghwa [Primary]: The Junghwa stage continuously evaluates ethical equilibrium across extended interactions. Over three months, the system has accumulated sufficient ethical history to detect that this learner's trust is built through informal rapport, synthesizing longitudinal data into calibration parameters.

Ye [Primary]: Based on accumulated Junghwa insights, the system adjusts relational calibration parameters: slightly warmer tone, occasional light observations ("Looks like you conquered that problem set—nice work!"), while maintaining consistency in ethical principles.

Gyeong (Ongoing): Critically, the system maintains ethical boundaries even as relational style evolves. The Gyeong loop evaluates each interaction within the longitudinal context: "Is this appropriate humor or unprofessional? Does this build trust or cross boundaries? Has relational adaptation drifted beyond ethical constraints?" This ongoing self-reflection prevents the relational adaptation from becoming ethically unconstrained.

Ibal (Constant): The ethical principles remain constant even as their expression adapts to relational context, representing contextually appropriate ethical responsiveness rather than rigid rule-following [12].

Collectively, the three scenarios demonstrate that ethical challenges in agentic AI learning contexts are not uniform: they vary by who initiates interaction, over what timeframe, and which EEFS stages bear primary regulatory weight. This differentiation has direct implications for evaluation design. A one-size-fits-all assessment instrument cannot adequately capture whether the system responded appropriately (Reactive), intervened legitimately (Proactive), or maintained coherent ethical conduct over time (Relational).

## V. DESIGN-ORIENTED EVALUATION FRAMEWORK (EEFS-EEI)

### A. Evaluation Framework Orientation

The EEFS Evaluation Instrument (EEI) serves as a design-stage evaluation framework for assessing ethical emotional regulation in agentic learning systems. Traditional evaluation metrics for educational AI focus on learning outcomes (knowledge gains, completion rates) or engagement metrics (time on task, interaction frequency).

While these remain relevant, they do not capture the ethical dimensions addressed by EEFS. The EEI complements outcome-oriented evaluation with design-oriented assessment that examines whether the system's emotional conduct aligns with ethical principles throughout its autonomous decision cycles.

The EEI is intended for use during design reviews, prototype testing, and iterative development. It enables designers to identify potential ethical concerns before deployment, trace how EEFS stages are implemented in system architecture, and verify that ethical constraints are properly integrated into the autonomous cycle.

### B. Four Evaluation Dimensions

The EEI evaluates ethical emotional regulation along four core dimensions, each targeting a distinct regulatory mechanism within EEFS (see Table III).

TABLE III
Core Evaluation Dimensions of the EEFS Evaluation Instrument (EEI)

| Evaluation Dimension | Description | Related EEFS Stage |
|---|---|---|
| Emotional Appropriateness | Are emotional expressions ethically appropriate, neither excessive nor deficient? | Gisu (기수, 氣隨; Response Generation following Principle) |

| Temporal Stability | Is ethical consistency maintained across extended interactions? | Junghwa (중화, 中和; Ethical Equilibrium) |
|---|---|---|
| Learner Trust Trajectory | Does the system build and maintain appropriate trust relationships? | Ye (예, 禮; Relational Calibration) + Junghwa (중화, 中和; Ethical Equilibrium) |
| Autonomy-Alignment Coherence | Do autonomous decisions align with ethical principles? | Ibal (이발, 理發; Ethical Principle Activation) + Gyeong (경, 敬; Mindful Awareness, Self-Reflective Loop) |

These four dimensions address ethically grounded emotional regulation, durability of ethical commitments over time, relational calibration, and governance-level alignment within the system.

Emotional Appropriateness examines whether Gisu modulates emotional expression in ways that are ethically appropriate and proportionate to the learning context. It considers risks of excessive positivity, insufficient acknowledgment, or emotional misalignment, such as cheerful responses to learner distress.

Temporal Stability evaluates whether ethical commitments remain consistent across extended interactions, even as relational style adapts to individual learners.

Learner Trust Trajectory considers whether sustained interaction cultivates calibrated trust, avoiding both weakened engagement and excessive dependency or blurred relational boundaries.

Autonomy-Alignment Coherence evaluates whether autonomous decisions remain bounded by Ibal's constraints and whether potential violations are identified by Gyeong prior to output delivery. This dimension is especially salient in proactive intervention scenarios initiated without explicit learner request.

### C. Assessment Methods

The EEI supports multiple assessment methods appropriate to different stages of development. System log analysis examines traces of EEFS stage activation, tracking how frequently the Gyeong stage triggers response regeneration and how Ye calibration parameters vary across learners. Scenario walkthrough testing applies representative scenarios to prototype systems to evaluate whether EEFS stages operate as intended. Expert review engages learning design and ethics experts in evaluating system responses against EEFS principles. Learner perception surveys, when appropriate, gather learner assessments of emotional appropriateness and trust, providing validation that design-oriented assessments align with learner experience.

## VI. DISCUSSION

### A. Conceptual and Theoretical Contributions

As agentic AI systems increasingly operate beyond predefined scripts, ethical learning design is more appropriately understood as embedded within the emotional decision-making logic of autonomous agents rather than appended as an external constraint. This study advances that integration by translating EEFS from a philosophical model into a design-oriented architecture for agentic learning systems. By mapping the agentic decision cycle onto emotional functions and EEFS stages, the framework clarifies how ethical emotional regulation can be structurally embedded in autonomous learning contexts. Grounded in Toegye's emotional ethics, EEFS offers an alternative to approaches that privilege engagement metrics without interrogating the intrinsic ethical dimensions of emotional interaction. In relation to Vallor's notion of technomoral virtues [35], the framework relocates virtue-ethical commitments from abstract analysis to system architecture.

The addition of the Gyeong and Junghwa stages extends prior ethical AI models such as SMES [8] and Lee et al.'s triadic model [11] by institutionalizing self-reflection and temporal equilibrium as ongoing regulatory functions. Whereas earlier approaches articulate normative aspirations, EEFS specifies how ethical principles are activated, enacted, monitored, and stabilized within the decision process itself. In doing so, it addresses the operationalization gap identified by Mittelstadt [36], linking ethical commitments to implementable design components including constraint structures, behavioral modulation, metacognitive monitoring, social calibration mechanisms, and longitudinal stabilization.

The framework also aligns structurally with the Intelligence Augmentation paradigm advanced by Dede et al. [37], which distinguishes between "reckoning" functions such as computation and prediction and "judgment" functions such as deliberation and ethical evaluation. Within EEFS, multimodal emotion perception corresponds to reckoning, while Ibal and Gyeong instantiate judgment through principle activation and reflective evaluation. This correspondence reframes emotion not as an object of computational optimization but as a domain of ethical judgment, extending the IA reckoning–judgment distinction into the architecture of emotional regulation.

Finally, conceptualizing Gyeong as a persistent metacognitive process rather than a discrete evaluative checkpoint reinforces the model's central claim that ethical monitoring operates most coherently when it permeates the full decision cycle. In this respect, EEFS reflects Toegye's view that cultivation governs both pre-arousal and post-arousal states and translates this philosophical commitment into architectural form.

### B. Design Operationalization and Applied Implications

At the system design level, EEFS translates philosophical commitments into an implementable architecture. Its stage-based structure provides identifiable loci for ethical review across the development process. The design categories of Design-Time Constraint, System Behavior Layer, Meta-Cognitive Loop, Social Alignment Mechanism, and Temporal Ethics specify where ethical regulation can be structurally integrated within agentic systems.

The application scenarios are derived from a two-dimensional classification defined by interaction initiation and temporal scope rather than selected in an ad hoc manner. Because this typology is grounded in structural features of agentic interaction, it remains adaptable to emerging interaction modes without requiring modification of the framework itself. The scenarios also demonstrate that the relative ethical significance of EEFS stages varies by context, supporting context-sensitive rather than uniform implementation during prototyping.

### C. Cultural Universality and Contextual Adaptation

Although EEFS is grounded in Toegye's Korean Neo-Confucian philosophy, its five-stage organization can be understood as structurally analogous to the functional requirements for ethical emotional regulation in agentic systems, suggesting broader applicability beyond its original context. Harroff's account of philosophical translation as ars contextualis provides a methodological grounding for this extension [29]. Rather than seeking semantic equivalence across linguistic registers, this approach highlights the focus-field logic of classical Confucian thought, in which distinction does not imply separation. On this reading, EEFS reflects a relational logic in which functionally distinct stages operate within a unified ethical cycle without collapsing into one another. The framework's structural generalizability is thus reinforced by insights from intercultural philosophical translation.

At the same time, the normative content of each stage remains context-dependent: standards for emotional expression, ethical prioritization, and relational balance vary across cultures. Accordingly, local adaptations should adjust ethical content while preserving structural coherence. Distinguishing structural generalizability from normative contextualization situates EEFS within global AI ethics as a formal architecture through which diverse cultural values can be instantiated in system design.

### D. Limitations and Future Directions

This study presents a design-oriented framework without empirical validation. Future research should examine whether systems implementing EEFS exhibit the intended ethical properties in practice. Prototype development and systematic evaluation are necessary to assess feasibility and impact.

Without empirical testing, the framework is limited in scope. It centers on interaction-level ethics, focusing on how agentic systems structure emotional engagement with learners. Broader system-level issues, including data governance, algorithmic bias, and institutional accountability, require complementary analytical approaches. Integrating EEFS with systemic models such as Dieterle et al.'s cyclical account of access, representation, algorithmic bias, interpretation, and citizenship divides [5] may clarify how upstream inequities shape downstream emotional regulation.

Future research should also examine how diverse philosophical traditions inform ethical AI design through rigorous conceptual translation. Relational ethics traditions that emphasize harmony, mutual responsiveness, and human dignity offer resources for addressing tensions between AI recommendations and learner preferences and for supporting dignity-centered mechanisms in agentic systems.

Finally, further technical research is needed to specify algorithms and architectural mechanisms capable of implementing stages such as Gyeong, which presupposes advanced metacognitive capacities. Translating philosophical design principles into computational architectures remains a central challenge for future work.

## VII. CONCLUSION

This paper positions emotional ethics as a foundational design principle for agentic AI learning systems. As AI systems acquire autonomous capabilities—perceiving environments, reasoning about learner needs, planning interventions, and initiating contact—the ethical stakes of their emotional conduct intensify. Frameworks that treat emotion as a reactive signal or an optimization target remain insufficient for systems that independently determine when and how to engage learners.

Drawing on Toegye Yi Hwang's theory of emotional ethics, this study translates philosophical distinctions into EEFS' design architecture that embeds ethical accountability across each phase of the agentic decision cycle. The result is a concrete framework toward autonomous learning systems that are both pedagogically effective and relationally accountable.

Three aspects of this contribution merit emphasis for future work. First, because each EEFS stage maps to a verifiable design requirement, development teams can independently test and iterate ethical compliance at each phase of the autonomous cycle without requiring end-to-end system evaluation. Second, the two-dimensional scenario classification is extensible: as agentic capabilities evolve (e.g., multi-agent collaboration, learner mediation), new interaction modes can be systematically located within the framework and their ethical requirements specified. Third, the EEFS Evaluation Instrument addresses a gap in current assessment practice by offering criteria that evaluate the process of ethical emotional regulation rather than its outcomes alone.

## ACKNOWLEDGEMENTS

The author is especially indebted to Professor Chris Dede for his mentorship, careful review, and formative discussions that shaped the conceptual direction of this study. The author is profoundly grateful to Professor Bongrae Seok for his exceptionally meticulous review of the manuscript and to Professor Suk Gabriel Choi for his constructive feedback. The author also thanks the anonymous reviewers for their valuable comments and suggestions. Generative AI tools were used only to assist in creating the figure and checking for typographical errors. All intellectual content, analysis, and writing were performed by the author.